\documentstyle[12pt,epsf]{article}
\newcommand{\be}{\begin{eqnarray}}
\newcommand{\ee}{\end{eqnarray}}

{
{

\def\0n{0\nu\beta\beta}

\begin{document}
\thispagestyle{empty}
\begin{center}
{\Large \bf 
Consequences of neutrinoless double beta decay and WMAP
} \\
\vspace{1.5cm}
{\large H.V. Klapdor-Kleingrothaus${}^1$ and U. Sarkar${}^{2}$}\\
\vspace{0.75cm}
{\sl ${}^1$Max-Planck-Institut f\"ur Kernphysik, P.O. 10 39 80}\\
{\sl D-69029 Heidelberg, Germany} \\
{\sl ${}^2$Physical Research Laboratory, Ahmedabad 380 009, India}
\end{center}

\vspace{2.5cm}

\begin{abstract}

Observation of the neutrinoless double beta decay ($\0n$) 
has established that there is lepton number violation in nature
and the neutrino masses are Majorana in nature. It also gives
the absolute mass of the neutrinos and discriminates between
different models of neutrino masses.  
The allowed amount of lepton number violation puts severe 
constraints on some possible new physics beyond the standard
model. The recent results from WMAP are consistent with the consequences  
of the neutrinoless double beta decay. They
improve some of these
constraints very marginally, which we shall summarise here. 
We mention 
the new physics which are not affected by WMAP and could make 
both these limits from the neutrinoless double beta decay and
WMAP consistent. 

\end{abstract}

\newpage

\section{Introduction}

During the past few years there have been several new results in
neutrino physics [1--6]. The atmospheric neutrino problem 
established that there is neutrino mass and that oscillations
occur between $\nu_\mu \to \nu_\tau$. This is
supported by K2K \cite{atm}. The solar neutrino problem
started quite some time back and experiments favored the 
large mixing angle MSW solution \cite{sol}. 
The KamLAND experiment has now confirmed 
that the large mixing angle MSW solution is the solution
of the solar neutrino problem \cite{sol1}. These experiments 
determine the two mass-squared differences and two mixing 
angles of the neutrino mass matrix. There 
is only an upper bound on the third mixing angle coming from the
reactor experiments \cite{chooz}. The neutrinoless double beta
decay tells us that the neutrinos are Majorana particles and also
provide us with the absolute mass \cite{9}. Recently the 
Wilkinson Microwave Anisotropy Probe (WMAP) has provided us with
a value for the total mass of the neutrinos and claims that
there are three generations of neutrinos \cite{wmap}. We shall 
restrict ourselves to only three generations of neutrinos.

A fit of all the data for atmospheric neutrinos and the K2K
gives \cite{atman}
\begin{equation}
\Delta m_{atm}^2 \simeq (1.8 - 4.0) \times 10^{-3} ~{\rm eV}^2,
~~~~ \sin^2 2 \theta_{atm} > 0.87 . \label{atmms}
\end{equation}
This oscillation is established to be between a $\nu_\mu$ and
a $\nu_\tau$ and the possibility of a sterile neutrino is ruled out.
A global fit to the results from the atmospheric, solar and the 
reactor neutrinos gives us two solutions 
for the solar neutrinos \cite{solan,solan2}. 
The allowed $3 \sigma$ region for the two solutions are 

\begin{center}
\begin{tabular}{ll}
LMA-I solution:&$5.1 \times 10^{-5} {\rm eV}^2 < 
\Delta m_{sol1}^2 < 9.7 \times 10^{-5} {\rm eV}^2 $\\
&\\
LMA-II solution:&$1.2 \times 10^{-4} {\rm eV}^2 <
\Delta m_{sol2}^2 < 1.9 \times 10^{-4} {\rm eV}^2 $\\
\end{tabular}
\end{center}
with mixing angle
$$ 0.29 < \tan^2 \theta_{sol} < 0.86 .$$
\noindent
For increasing $\sin^2 \theta_{13}$ the LMA solutions shrink 
and eventually disappear, which requires $\sin^2 \theta_{13} < 0.04$
\cite{solan}. Again, a sterile neutrino solution is completely
ruled for the solar neutrinos. 

An analysis of the Heidelberg-Moscow data yields a half-life for
the neutrinoless double beta decay experiment of \cite{9}
$$ T_{1/2}^{0 \nu} = (0.7 - 18.3) \times 10^{25} ~~{\rm y}
~~~~~~~{\rm at ~95 ~\% ~~c.l.} $$ with a best value of $1.5 \times
10^{25}$ y.
The signal for the neutrinoless double beta decay amounts to
an effective Majorana neutrino mass of the electron neutrino in
the range of \cite{9}
\begin{equation} 
\langle m \rangle = (0.11 - 0.56) ~ {\rm eV} ~~~~~~~~ {\rm at ~ 95\% ~~c.l.}
\end{equation}
with a best value of $0.39$ eV (with nuclear matrix element of 
reference \cite{numat}). A weaker bound for the neutrino mass of
\begin{equation}
\langle m \rangle = ( 0.05 - 0.86) ~ {\rm eV} ~~~~~~~~ {\rm at ~ 95\% ~~c.l.}
\end{equation}
is deduced,
when a $\pm 50\%$ uncertainty in this nuclear matrix element is
allowed (for details see \cite{9}).  
We shall not discuss the bounds on the neutrinoless double 
beta decay from another 
analysis \cite{vogel1} that uses a 15-years old nuclear matrix element. 
In that calculation of the nuclear matrix elements they 
did not include a realistic nucleon-nucleon
interaction, which has been included by all other calculations of the
nuclear matrix elements over the last 15 years. This matrix element has 
been ruled out by the WMAP result completely \cite{mur}.

On the basis of the most recent result from WMAP it is claimed that
there are three generations of neutrinos and that the LSND result is
ruled out \cite{mur}. A limit on 
the total neutrino masses of 
\begin{equation}
m_s = \sum m_\nu < 0.69 ~{\rm eV~~~~~~~ at~ } 95\%~~ {\rm c.l.},
\end{equation}
is given by the analysis of ref. \cite{wmap}. It has been shown,
however, that this limit may not be very realistic. 
Another analysis shows that this limit on the total mass should be \cite{hann}
\begin{equation}
m_s = \sum m_\nu < 1.0 ~{\rm eV~~~~~~~ at }~ 95\% ~~{\rm c.l.} 
\end{equation}
The latter analysis also
shows, however, that four generations of neutrinos are still allowed and in the
case of four generations the limit on the total mass is 
increased to $1.38$ eV. If there is
a fourth neutrino with very small mass, then the limit on the total mass
of the three neutrinos is further weakened and there is essentially no
constraint on the neutrino masses. 
In our analysis we comment on these possible values. 

\section{Constraints on models of neutrino masses}

There are several consequences of the neutrinoless double beta
decay. Following the announcement of the positive evidence for
the process, lots of activity started (see {\it e.g.} \cite{ndbdh}). 
We concentrate on a few aspects only.
We first discuss the different scenarios of neutrino masses. We
parametrize the neutrino mass eigenvalues as
\begin{equation}
\Delta m_{sol}^2 = |\Delta m_{12}^2| ~~~~ {\rm and} ~~~~
\Delta m_{atm}^2 = |\Delta m_{23}^2|
\end{equation}
where $\Delta m_{12}^2 = |m_2|^2 - |m_1|^2$ and 
$\Delta m_{23}^2 = m_3^2 - |m_2|^2$. For the mixing angle in 
the atmospheric neutrinos we assume maximal mixing, so that
$\sin^2 2 \theta_{atm} = \sin^2 2 \theta_{23} = 1$, {\it i.e.},
$\sin \theta_{23} = \cos \theta_{23} = 1/\sqrt{2}$. For solar
neutrinos we allow the entire allowed range and write
$\cos \theta_{12} = c$ and $\sin \theta_{12} = s$, with
$ 0.54 < s/c < 0.93 $ or $0.48 < s < 0.68$ and a best value
of $s = 0.55$. Taking the limit on the third mixing
angle, we assume, $\cos \theta_{13} \approx 1$ and $\sin
\theta_{13} = u < 0.2$. In our analysis we do not include
CP violating phases and hence the mixing matrix may be parametrized
as
\begin{equation}
U = \pmatrix{ c & s & u \cr -(s + c u)/\sqrt{2} & (c - s u)/
\sqrt{2} & 1/\sqrt{2} \cr (s - c u)/\sqrt{2} & -(c + s u)/
\sqrt{2} & 1/\sqrt{2} } .
\end{equation}
As noted in earlier references \cite{ndbold}, inclusion of CP violation does
not affect most of the conclusions regarding the bounds on the
neutrino mass. 
The neutrinoless double beta decay bound then implies
\begin{equation}
m_1 c^2 + m_2 s^2 + m_3 u^2 = \langle m \rangle,
\end{equation}
and the WMAP implies
\begin{equation}
|m_1|  + |m_2|  + |m_3|  < m_s.
\end{equation}
In some models it is possible to have neutrinoless double beta
decay mediated not be exchange of a massive Majorana neutrino, but
of some exotic particles \cite{ma,uehara}. In that case it would
make no sense to compare limits from WMAP or corresponding experiment
with neutrinoless double beta decay.

In general, it may be possible to classify the neutrino masses in
four classes:

{\bf Hierarchical:} 

This is the most natural choice
for neutrino masses, where all three neutrino masses are different
and hierarchical ($m_1 < m_2 < m_3$) 
similar to the mass hierarchy of the charged fermions. This implies
$m_3 = m_{atm} = \sqrt{\Delta m_{atm}^2}$,
$m_2 = m_{sol} = \sqrt{\Delta m_{sol}^2}$ and 
$ m_1 < m_{sol}$. The WMAP bound then becomes,
$|m_3| < m_s$, since the other two masses are too small. 
If we use the highest value of $m_1$ to be $m_{sol}$ and 
use the WMAP constraint along with CHOOZ and solar neutrino results,
then the contribution to the effective mass appearing in the neutrinoless
double beta decay becomes $$ \langle m \rangle ~ < m_{sol} + m_3 u^2 <
m_{sol} + m_s u^2 < 0.027 ~ {\rm eV} $$ for $m_s = 0.69$. 
Thus the WMAP constraint sets a stronger limit compared to 
the neutrinoless double beta decay. But it still allows the
hierarchical solution with solar and atmospheric neutrino
solutions, whereas the neutrinoless double beta decay result
rules out this solution. This is true also for a higher 
value of the WMAP constraint of $m_s = 1.0$, where $ \langle m \rangle ~ < 
0.04$ eV. However, when the atmospheric
neutrino constraint is considered $m_3 = m_{atm}$, the WMAP condition 
is trivially satisfied. In this case 
the contribution to the neutrinoless double beta decay is further
reduced and becomes too small to explain the present bounds
$\langle m \rangle 
~< m_{sol} + m_{atm} u^2 < 0.016$ eV. We considered the maximum value
for $m_1 \sim m_{sol}$ and the largest values of all the parameters.
There is no lower limit on the contribution to $\langle m \rangle $
in this case.
In brief, neutrinoless double beta decay does not allow the hierarchical 
neutrino mass matrix. The WMAP limit cannot yield such a strong statement.

{\bf Degenerate:} 

This is the most interesting solution at present,
which is allowed by all the experiments. Here one assumes 
$m_1 \approx m_2 \approx m_3 \approx m_0$, where $m_0$ is the 
overall mass and the mass squared differences are as required by
the solar and atmospheric neutrinos. 
The WMAP bound then implies $m_0 < m_s/3$. The
neutrinoless double beta decay constraint now implies $\langle m \rangle \leq
m_0 < m_s/3$ and hence part of the allowed region is ruled out by the WMAP
result. For $m_s = 0.69$ eV, this implies $\langle m \rangle < 0.23$ eV and the 
best fit value for the neutrinoless double beta decay is not within
the range. On the other hand, for
the other more realistic 
limits of $m_s = 1.0$ eV and $1.38$ eV, we get $\langle m \rangle 
~< 0.33$ eV
and $0.46$ eV respectively, which does not conflict with the
best fit value of the neutrinoless double beta decay. For the
degenerate solution the lowest
contribution to the neutrinoless double beta decay would correspond to
$m_0 = m_{atm}$ and comes out
to be $\langle m \rangle ~> m_{atm} (c^2 - s^2 -u^2) > 0.001$ eV.

{\bf Inverted Hierarchical:} 

In this case one considers
that two of the neutrinos are degenerate and heavier than the third
one, $m_1 \approx m_2 > m_3$. The $\Delta m_{12}^2$ is very small
and can explain the solar neutrino
problem, while $m_1 \sim m_{atm}$ so that $\Delta m_{23}^2$ 
can explain the atmospheric neutrino problem. The WMAP bound now
implies that $m_1 = (m_1 + m_2)/2 < m_s/2 = 0.345$ eV for $m_s = 0.69$.
In this case the effective mass for the neutrinoless double beta decay
becomes $$\langle m \rangle = m_1 c^2 + m_2 s^2 < m_1 < m_s/2 < 0.345 
~ {\rm eV}. $$ for $m_s = 0.69$. For $m_s = 1.0$ the bound is $0.5$ eV.
The WMAP result does not allow part of the allowed region 
of the neutrinoless double beta decay, which is anyway not allowed
by the atmospheric and solar neutrinos. 
The atmospheric neutrino solution requires $m_1 \sim m_{atm}$,
so the contribution to the 
neutrinoless double beta decay becomes $\langle m \rangle ~< m_1 \sim 
m_{atm} < 0.06$ eV. This value is still marginally allowed, when allowing for
a $\pm 50\%$ uncertainty in the matrix element of \cite{9}. The
lowest allowed value now corresponds to $\langle m \rangle ~>
m_1 (c^2 - s^2) = 0.003$ eV.

{\bf Partially Degenerate:} 

This scenario is ruled out by
neutrinoless double beta decay, but still allowed by atmospheric and 
solar neutrinos and also WMAP. There are two degenerate neutrinos whose
mass difference squared $\Delta m_{12}^2$ solve the solar neutrino 
problem and the third neutrino is heavier $m_1 \approx m_2 < m_3$. 
A solution to the atmospheric neutrinos requires $m_3 = m_{atm}$.
The main difference between the hierarchical and this scenario is
that, in this case $m_1 \approx m_2 > m_{sol}$.
To distinguish this solution from the degenerate scenario, we consider
$m_1 < m_{atm} = 0.4$ eV. 
The WMAP constraint now implies $m_3 < m_s = 0.69$ eV. In this case
we can have $m_1 \approx m_2$ to be close to but less than $m_3 < m_s$.
So, the effective mass for the neutrinoless double beta decay becomes
$$ \langle m \rangle = m_1 c^2 + m_2 s^2 + m_s u^2 < m_1 < m_s
= 0.69 ~ {\rm eV}. $$ For $m_s = 1.0$ this bound is also high as
shown in the figure. If we now include the atmospheric neutrino result, 
the WMAP constraint is satisfied. The
neutrinoless double beta decay contribution now becomes
$\langle m \rangle = m_1 c^2 + m_2 s^2 < m_1 < m_{atm}$. 
The highest value of $m_1$ and $m_2$ correspond to $m_1 \approx 
m_{atm}$, which is included in the degenerate solution. So, if
we restrict the partially degenerate solution to $m_1 < m_{atm}$, then
the contribution to the neutrinoless double beta decay becomes
$\langle m \rangle < 0.04$ eV and hence the neutrinoless double
beta decay does not allow this solution, even when we relax the
uncertainty in the nuclear matrix elements. 

\begin{figure}[htb]
\vskip 0in
\epsfxsize=85mm
\centerline{\epsfbox{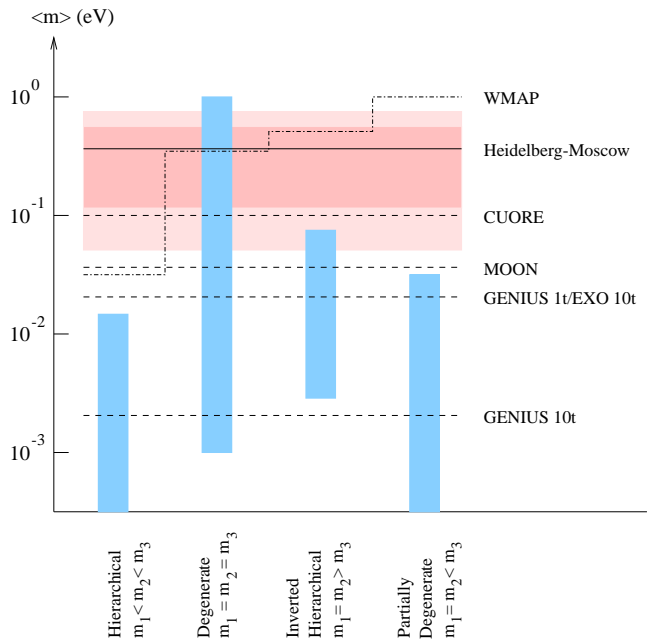}}
\vskip 0in
\caption{
Contributions in different models to the
neutrinoless double beta decay.
The present result is given by the dark shaded region
(the solid line denoting the best value and the light
shaded region allowing $\pm 50\%$ uncertainty in the nuclear
matrix element). The WMAP line is plotted for $\sum m_\nu < 1.0$ eV,
although it could be even weaker ({\it i.e.}, the
line running higher) as mentioned in the text. 
Future sensitivity that might be reached for the CUORE \cite{cuore},
MOON \cite{moon} and the one ton and ten tons GENIUS
\cite{genius} are given for comparison.
\label{ndbchfg}}
\end{figure}

All the four scenarios we mentioned are solutions of the solar
and atmospheric neutrino problems. All these solutions are also allowed
by the WMAP result, except in the degenerate case when a small part of the
solution is not allowed. On the other
hand, the neutrinoless double beta decay does not allow the 
hierarchical and the partially degenerate solutions. The inverted
hierarchical solution is only marginally allowed by the neutrinoless
double beta decay when an extra $\pm 50\%$ 
uncertainty in the nuclear matrix element is permitted. Only the 
degenerate solution is allowed by the entire neutrinoless double beta 
decay range, but a small part of the allowed region is 
ruled out by the WMAP result. 
This is shown in figure \ref{ndbchfg}. In the figure we 
considered the more recent analysis of the WMAP result \cite{hann}
and used the bound $m_s < 1.0$ eV.

\section{Constraints on lepton number violation}

The neutrinoless double beta decay process could be triggered also
by exchange of other particles, than massive neutrinos \cite{rev}.
In this sense a deduced effective mass is - though the most natural 
explanation - strictly only an upper limit. The measured half-life (or 
its lower bound) can thus be used to deduce limits for other beyond 
standard model physics and other lepton number violating interactions. 

The WMAP constraint depends on which value we consider. If
we consider the value quoted in the original paper, then it 
implies a slightly improved bound on some of the lepton number violating 
processes. However, if the weaker bounds are considered, then
it does not improve any of the constraints compared to the present
bounds on the neutrinoless double beta decay. In the rest of the
analysis we shall consider the value $\sum m_\nu < 1.0$ eV, so that
combining with the neutrinoless double beta decay we get a 
limit on the effective mass of $\langle m \rangle < 0.33$ eV. 

If there are heavy right-handed neutrinos, which have small mixing
with the left-handed neutrinos, then they can enter the neutrinoless
double beta decay processes and contribute to the effective mass. The
present bound on the lifetime of the neutrinoless double beta decay
would then give a constraint \cite{18} 
\begin{equation}
M_N > 6 \times 10^7 ~ {\rm GeV} .
\end{equation}
From the same analysis the bound on the
the right-handed $W$ boson comes out to be 
\begin{equation} 
m_{W_R} \geq 1.2 ~\left( {M_N \over 1 ~{\rm TeV}} \right)^{-1/4}
~ {\rm TeV}.
\end{equation}
With some reasonable theoretical input, this may be translated to 
an absolute lower bound of $m_{W_R} > 1.2 $ TeV \cite{18}. Including 
WMAP constraints this bound will be improved to  $m_{W_R} > 1.5 $ TeV. 

Using the lifetime of neutrinoless double beta decay, the probability
for the discovery of the inverse beta decay process $e^- e^- \to W^- W^-$ 
at NLC could be constrained. 
The present value \cite{9} can be achieved at a future linear collider NLC 
when it reaches a center of mass energy of 2 TeV \cite{19,19a}.
There is hardly any change in the analysis when the WMAP result is
considered. 

If there are 
Higgs scalar bilinears, which couple to the usual quarks and
leptons, they can also allow for neutrinoless double beta decay. 
However, in this case it is possible that the contribution of
these scalars to the neutrino mass is negligible \cite{ma}. 
Then these scalars could allow the neutrinoless double beta
decay as claimed, but the neutrino mass will be much smaller
and hence there will not be any constraint from the WMAP result.
In other words, in these scenarios it is possible to satisfy even
the stronger WMAP bounds simultaneously making it consistent with 
the neutrinoless double beta decay. 

Although the dileptons were first considered \cite{21} in 
connection with the left-right symmetric model, no significant bound
is possible on this scalar. For the leptoquarks the bound is
better. If the $X-$type leptoquarks ($SU(2)_L$ singlets) mix with the $Y-$type
leptoquarks ($SU(2)_L$ doublets), 
then they can give an effective operator $u \bar \nu \bar d
\bar l$ that generates a diagram contributing to the
neutrinoless double beta decay involving the leptoquarks \cite{22}. A 
mixing between these two leptoquarks could take place only after
the electroweak symmetry breaking, if both these leptoquarks
couple to the usual standard model Higgs doublet $\phi$. In that
case a coupling $\phi X Y$ will induce a mixing of $X$ with $Y$
when $\phi$ acquires a $vev$. It was
noticed \cite{22} that in the leptoquark mediated case, there is a huge
enhancement factor of ${<q> \over m_\nu} \sim 10^8 ~(1 {\rm eV}/m_\nu)$,
where $m_\nu$ is the effective neutrino mass entering the 
neutrinoless double beta decay contribution, and $<q>$ is the 
Fermi momentum of a nucleon inside a nucleus, which is about 
$200-300$  MeV. For a leptoquark with mass of the order of 100 GeV, the
effective coupling constant (including the mixing contribution)
comes out to be about $10^{-9}$. The WMAP result does not modify
this bound.

There are other exotic scalar bilinears, which can also mediate
the neutrinoless double beta decay \cite{23}. From the present
allowed range of the neutrinoless double beta decay it is thus
possible to put severe constraints on these scalars. If we assume
a common mass for these scalars of about 100 GeV and that the
self interaction of these scalars is of the order of 1, a strong
bound on the effective coupling of the scalars to ordinary
fermions becomes $f < 10^{-7}$. In other words, if the couplings 
are assumed to be of the order of 1, there is a lower bound on the
masses of these scalars to be of the order of $10^4$ GeV.
In general, a constraint on the ratio of the masses to their coupling to first
generation fermions for all the exotic scalar bilinears could be given
\cite{23}, which is comparable to the bounds from other processes like 
$K^\circ - \overline{K^\circ}$ oscillations,
$B^\circ - \overline{B^\circ}$ mixing, $D^\circ - \overline{D^\circ}$ 
mixing, proton decay or $n - \bar n$ oscillations. Again, these bounds
are not modified by the WMAP result. Only in some cases, when the
exotic particles also give a large neutrino mass, the present bound
is improved very marginally by the WMAP result. 
Considering the uncertainty in the matrix elements for
these processes, these bounds are not worth mentioning. 

The indirect bounds discussed also constrain 
some of the possibilities of composite particles. For example,
if a neutrino is a composite particle, the most severe constraint
comes from the neutrinoless double beta decay \cite{24},
\begin{equation}
|f| \leq 3.9 {\Lambda_c \over 1~ {\rm TeV}} \left({M_N \over
1 ~ {\rm TeV}}\right)^{1/2}
\end{equation}
where $\Lambda_c$ is the compositeness scale, $M_N$ is the
mass of the heavy excited neutrino and $f$ is the dimensionless
coupling constant. This bound is also not affected by the WMAP
result. 

It is also possible to constrain several parameters of
supersymmetric theories. Recently one analysis claimed that the
WMAP result improves the existing bounds on these parameters
by about one order of magnitude \cite{unreas},
but we find this claim unreasonable. If we consider the 
bound from the neutrino mass from the neutrinoless double beta decay,
then there is hardly any change in this bound 
after including the WMAP result. For the bounds available in the
literature for the supersymmetric theories coming from the 
neutrinoless double beta decay see for example, ref. \cite{27,27a}. Given
the uncertainty in the calculations of these models, this change 
in the number is negligible. We do not present here these
unchanged numbers.

The bounds on the sneutrino-antisneutrino oscillation is also
unchanged compared to the earlier bounds \cite{25}. There are also
bounds on the scale of extra dimensions in models in which 
mini black holes generate neutrino mass \cite{ue}. That limit
is modified very marginally. The allowed textures of the 
neutrino masses are also constrained by the neutrinoless double
beta decay, which are not affected by the WMAP results \cite{kkus}. 

\section{Summary}

We studied the consequences of the neutrinoless double beta decay
and the WMAP results. Models of neutrino masses are severely 
constrained by the neutrinoless double beta decay, while the WMAP
result may only eliminate a small part of the region allowed by the
neutrinoless double beta decay. WMAP constraints become the strongest
in the case of hierarchical solution, but it is not ruled out. On
the other hand, the
hierarchical solution is ruled out by the neutrinoless double 
beta decay. The degenerate solution is most favored by the neutrinoless
double beta decay and only a small part of the allowed region is constrained by
the WMAP result. WMAP bounds become weaker for the inverted hierarchical 
and partially degenerate solutions, whereas neutrinoless double beta decay
can only marginally allow the inverted hierarchical solution when an
extra $\pm 50\%$ uncertainty in the nuclear matrix element is allowed.
The constraints on the lepton number
violating processes are severely constrained by the neutrinoless double
beta decay, while only a few of these constraints are negligibly 
modified by the WMAP result. 

{\bf Acknowledgement} One of us (US) thanks 
the Max-Planck-Institut f\"ur Kernphysik for hospitality.

\end{document}